\documentclass[twocolumn,showpacs,preprintnumbers,amsmath,amssymb]{revtex4}

\usepackage{graphicx}
\usepackage{dcolumn}
\usepackage{bm}

\begin{document}


\title{Thermodynamic Emergence of a Brownian Motor}
\author{Alexander Feigel}
\email{sasha@phys.huji.ac.il}
\affiliation{Racah Inst. of Physics, Hebrew University of Jerusalem, Israel}
\author{Asaf Rozen}
\affiliation{Physics Department, Weizmann Inst. of Science, Rehovot Israel}
\date{\today}
\begin{abstract}
Human-created engines and evolutionarily optimized molecular motors exhibit sophisticated design in order to harvest chemical or thermal energy for generating unidirectional motion. The complexity of these motors makes their random emergence unlikely. Molecules capable of locomotion, however, seem to be essential to the creation of the first self-replicator and initiation of Darwinian evolution, posing a question of the physical mechanism that can facilitate emergence of directed motion in an isotropic environment. Here we show a universal thermodynamic mechanism for spontaneous emergence of motor abilities in a mechanical system. A non-equilibrium system with multiple degrees of freedom develops symmetry breaking that favors rectification of environmental thermal fluctuations. The corresponding velocities and its fluctuations are calculated. Homochirality of living matter is explained as chirality breaking resulting from the emergence of a motor. Universality of the results provides a general extension of the Onsager relations to the non-linear regime.
\end{abstract}


\maketitle

\begin{figure}
\begin{center}
    \begin{tabular}{c c}
\multicolumn{1}{l}{{\bf\sf A}} & \multicolumn{1}{l}{{\bf\sf B}}\\
\resizebox{0.25\textwidth}{!}{\includegraphics{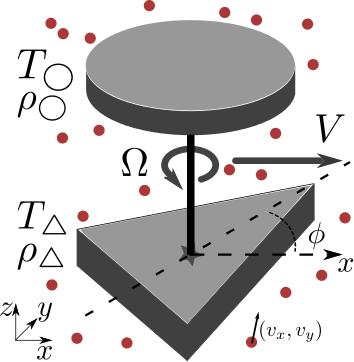}} & \resizebox{0.25\textwidth}{!}{\includegraphics{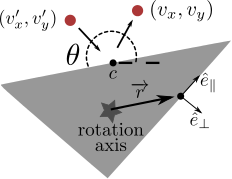}}\\
    \end{tabular}
\caption{Brownian Motor with translation and rotation degrees of freedom. (A)  The motor is an extension of the single degree of freedom Triangulita model\cite{VandenBroeck2004}. It consists of triangle and symmetric (e.g. oval) parts that are immersed into two different thermal bathes. The bathes consist of ideal gases with particles of mass $m$, temperatures $T_{\bigcirc}$, $T_{\bigtriangleup}$ and densities $\rho_{\bigcirc}$, $\rho_{\bigtriangleup}$ respectively. Velocities of the gas particles $(v_{x},v_{y})$ fit Maxwell distribution. The Motor possesses velocity $V=\partial x /\partial t$, along $x$ coordinate and frequency $\Omega=\partial\phi /\partial t$, where $\phi$ is the angle between the axis of triangle and axis $x$. Velocity $V$ takes it maximum absolute values when triangle is directed along axis $x$, $\phi=0,\pi$. $V$ vanishes due to symmetry at $\phi=\pi/2,-\pi/2$. (B). The motor interacts with the thermal bathes by scattering of the gas particles. The scattering depend on the geometry of the body, such as an angle between axis $x$ and tangential of the surface. Final change in momentum and in energy of the motors requires integration of the interactions at all points $c$ along the surface of the body.}
\label{fig1}
\end{center}
\end{figure}

Thermal fluctuations of the environment cause random motion of microscopic particles. This Brownian motion can be rectified: for instance, two connected particles exhibit unidirectional motion given that their local environments possess different temperatures\cite{VandenBroeck2004}, see Fig \ref{fig1}. Motion emerges in this non-equilibrium system if, in addition to the temperature difference, at least one of the particles is asymmetric. One can then ask a question: if the particles can be in either symmetric or asymmetric state, will thermodynamic forces favor a transition to the asymmetric state and, therefore, the emergence of motion? Answering this question may shed light on a major biological and evolutionary puzzle, since rectified Brownian motion is believed to provide a major mechanism for directed molecular motion in living cells\cite{Hoffmann2012}.  

Rectification of thermal fluctuations was first proposed by M. Smoluchowski\cite{Smoluchowski1912} and popularized by R. Feynman\cite{Feynman2011} as the Brownian ratchet. The first theoretical treatment of the relevant phenomena in semiconductors\cite{BUG1987,BUTTIKER1987,VANKAMPEN1988} was followed by the seminal works\cite{Ajdari1992,Magnasco1993,Magnasco1994,DOERING1994,MILLONAS1994,BARTUSSEK1994,AJDARI1994a,PROST1994,ASTUMIAN1994,Peskin1993,HONDOU1994} motivated by an analogy between Brownian ratchets and the mechanical motion of proteins in living matter (for review, see\cite{Hanggi2009,Reimann2002,Seifert2012}). Motion is essential for transcribing and translating nucleotide sequences into functional proteins; it also plays key roles in self-replication and cellular motility. Recent experiments have indicated the presence of multiple ratchet-like processes in the proteins responsible for these cellular functions\cite{Liu2014}.
 
Models of the Brownian motors fit Onsager formalism $\partial x_{i}/\partial t=L_{ij}F_{j}$, where $x_{i}$ are degrees of freedom and $F_{j}$ are the corresponding mechanical or thermodynamic forces\cite{Onsager1931,Julicher1997,Gaspard2013}. Thus emergence of rectification is analogous to emergence of Onsager coefficients $L_{12}\neq 0$ in relations:
\begin{eqnarray}
  \label{eq:3}
  \frac{\partial x}{\partial t} = L_{11}F_{X}+L_{12}\frac{\Delta T}{T^{2}},\;\;
  \frac{\partial Q}{\partial t} = L_{21}F_{X}+L_{22}\frac{\Delta T}{T^{2}},
\end{eqnarray}
where $x$ is translation coordinate, $F_{j}$ is the corresponding mechanical force, $Q$ is thermal flux and $\Delta T/T^{2}$ is the corresponding thermodynamic force. This set of equation describes a motor that generates motion along axis $x$ using thermal flux $Q$ between two thermal bathes with temperature difference $\Delta T$. 

In general, emergence of Onsager coefficients can be explained by the transition of a system to the corresponding state due to the minimization of energy or a random event. These explanations, however, are debatable when applied in the Origin of Life. The simplest modules of living systems, e.g. molecular motors\cite{Vale2000}, seem to be too complex to emerge by chance. In addition to that, emergence by chance or by energy minimization predicts equal probability for either L or R chiralities of bio-molecules. Al living matter, on the contrary, is composed of L chiral aminoacids and R chiral sugars\cite{Guijarro2008,Barron2008}.

Emergence of Onsager coefficients $L_{ij}$ may occur in the case of non-linear Onsager relations, $\partial x_{i}/\partial t = L_{ij}(x_{k})F_{j}$, where coefficients $L_{ij}$ depend on the degrees of freedom $x_{k}$. In this case, multiple steady states may exist. Development of motion corresponds to convergence to a steady state with finite Onsager coefficient for the rectification of thermal fluctuations, e.g. $L_{12}\neq 0$ in (\ref{eq:3}). Linear Onsager relation possesses many universal properties such as symmetry $L_{ij}=L_{ji}$ and minimum entropy production at the steady states $\partial x_{i}/\partial t = 0$. Recently, the limits on the Onsager coefficients\cite{Brandner2013}, on the fluctuations\cite{Gingrich2016,Pietzonka2015,Bodineau2004} of the fluxes $\partial x_{i}/\partial t$ and on the corresponding dissipation were discussed in light of the general universal properties of non equilibrium systems, such as Jarzynski equality\cite{Jarzynski1997,Crooks1999}. To the best of our knowledge, the universal extension of Onsager relations to non-linear regime remains an open question.

In this Letter we demonstrate a universal mechanism for emergence of rectified Brownian motion in a mechanical system out of thermal equilibrium. This mechanism favors motion with specific chirality. The analysis fit general mechanical system with $N$ degrees of freedom that are coupled to arbitrary amount of thermal bathes. The corresponding non-linear form of Onsager relation is, therefore, universal. The emerging fluxes and their fluctuations are calculated analytically as functions of geometric factors that define linear viscous coefficients at equilibrium. The implications include an explanation for the emergence of molecular motors with specific chirality and novel analytic tools for analysis of systems out of thermal equilibrium. The discussion includes implication of the findings on the Origin of Life question.

Following\cite{VandenBroeck2004} and\cite{Broek2008}, consider a dumbbell-like macroscopic body that consist of symmetric and asymmetric two dimensional parts, see \ref{fig1}. These parts are rigidly connected to each other by a thin axis. Each part is in contact with different thermal bath: infinite reservoir of ideal gas with identical particles of mass $m$ at density $\rho_{\bigtriangleup}$ and temperature $T_{\bigtriangleup}$ around asymmetric part and density $\rho_{\bigcirc}$ and temperature $T_{\bigcirc}$ around asymmetric part. The body possesses two degrees of freedom: First, it can move along $x$ axis with velocity $V$. Second, it can rotate around $z$ axis with frequency $\Omega$. The rotation coordinate is $\phi$. The kinetic energy of the body, therefore, is $MV^{2}/2+I\Omega^{2}/2$, where $M$ is the mass and $I$ is the momentum. Both degrees of freedom are translation invariant because no potential is present. Gas particle move only in $(x,y)$ plane with velocities $(v_{x},v_{y})$ according to Maxwell distribution.

To calculate velocities $V$ and $\Omega$ as a function of the corresponding coordinates $x$ and $\phi$, the first task is to derive transition probabilities $W(V,V')$ and $W(\Omega,\Omega')$ from velocity $V'\rightarrow V$ and frequency $\Omega'\rightarrow \Omega$ due to scattering of the gas particles. Single particle may scatter at any point $c$ along perimeter of the motor. The parameters that are required to calculate the velocities of the motor after the scattering are angle $\theta$ between $x$ axis and the surface together with radius $\overrightarrow{r}$ of the point of interaction relative to the axis of rotation. The scattering is assumed to be elastic. Consequently, total energy and momentum conserve together with the momentum of the gas particle along the surface of the body at the scattering point $c$. Transition probability rates $W$ are averaged over all interaction in all points $c$ and all velocities of the gas particles.

Single degree of freedom cases\cite{VandenBroeck2004,Broek2008,Meurs2004}, only translation motion is present while rotation coordinate is fixed $\phi= const,\Omega=0$  or vise verse body rotates at $x=const,V=0$, possesses strikingly similar description. In both cases transition probability rates $W$ are:
\begin{eqnarray}
  \label{eq:30757}
  &&W(V,\Delta V)=\frac{1}{4}\sum_{i}S_{i}\rho_{i}\sqrt{\frac{m}{2\pi T_{i}}} \oint_{c}  \\\nonumber
&&\left |\Delta V\right |H(\Delta V \Gamma_{V,i}(c))\Gamma_{V,i}^{2}(c)\left (\frac{M}{m\Gamma_{V,i}^{2}(c)}+1\right )^{2}\\\nonumber
&&\exp\left [-\frac{m\Gamma_{V,i}^{2}(c)\left ( V+\frac{1}{2}\left[ \Delta V\left (\frac{M}{m\Gamma_{V,i}^{2}(c)}+1\right )\right ]\right )^{2}}{2T_{i}}\right ]
\label{eq:31}    
\end{eqnarray}
where $\Delta V = V-V'$, index $i$ goes over the thermal bathes. Geometric coefficient $\Gamma_{V}=\sin\theta$ also defines linear viscous coefficient at thermal equilibrium:
\begin{eqnarray}
  \label{eq:337587}
  \gamma_{i}=4S_{i}\rho_{i}\sqrt{\frac{mT_{i}}{2\pi}}\oint_{c}\Gamma_{V,i}^{2}(c)
\end{eqnarray}
The same expression holds for rotation $W(V,\Delta V)\rightarrow W(\Omega,\Delta\Omega)$ with substitutions $M\rightarrow I$ and $\Gamma_{V,i}\rightarrow\Gamma_{\Omega,i}$, where $\Gamma_{\Omega}=r_{x}\cos\theta+r_{y}\sin\theta$.

Single degree of freedom analysis\cite{VandenBroeck2004,Broek2008,Meurs2004} predicts motion as a function of the temperature difference between the thermal bathes and the contour integral of the third degree of the geometric coefficient:
\begin{eqnarray}
  \label{eq:3785554}
  <V> = \sqrt{\frac{m}{M}}\sqrt{\frac{\pi }{8M}}\frac{\sum_{i}S_{i}\rho_{i}(T_{i}-T^{eff}_{V}{})\oint_{c}\Gamma^{3}_{V,i}(c)}{\sum_{i}S_{i}\rho_{i}T^{1/2}_{i}\oint_{c}\Gamma_{V,i}^{2}(c)},
\end{eqnarray}
where $T^{eff}_{V}$ is:
\begin{eqnarray}
  \label{eq:11}
 T^{eff}_{V}=\sum S_{i}\rho_{i}T^{3/2}_{i}\oint\Gamma^{2}_{V,i}/\sum S_{i}\rho_{i}T^{1/2}_{i}\oint\Gamma^{2}_{V,i}. 
\end{eqnarray}
Expression for rotation is obtained by substitutions  $V\rightarrow\Omega$, $M\rightarrow I$ and $\Gamma_{V,i}\rightarrow\Gamma_{\Omega,i}$. Linear expansion of (\ref{eq:3785554}) near thermal equilibrium (the temperatures of the thermal bathes are identical) corresponds to (\ref{eq:3}), $<V>\propto L_{12}\Delta T/T^{2}$.

In the case of single degree of freedom, motion (\ref{eq:3785554}) vanishes if both parts of the body are symmetric relative to the axis $x$. In this case $\oint\Gamma_{V}^{3}=0$. In the case of triangle asymmetric part, there is no translation motion if the symmetry axis of triangle is perpendicular to axis $x$. Axis of the triangle which is parallel to the axis $x$ corresponds to the maximum absolute velocity for a given temperature difference between the thermal bathes. Rectification of rotation vanishes $<\Omega>=0$, if no translation motion is allowed and asymmetric part possesses triangle shape, because $\oint\Gamma^{n}_{\Omega}=0$ disregard of triangle orientation.

\begin{figure}
\begin{center}
    \begin{tabular}{c}
\multicolumn{1}{l}{{\bf\sf A}}\\
\resizebox{0.5\textwidth}{!}{\includegraphics{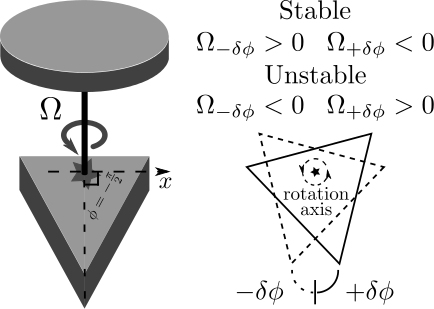}}\\
\multicolumn{1}{l}{{\bf\sf B}}\\
\resizebox{0.5\textwidth}{!}{\includegraphics{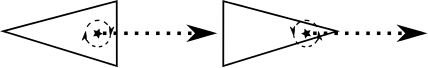}}\\
    \end{tabular}
\caption{Stability of the rectified motion. (A) Emergence of the motion corresponds to an instability of the symmetric configurations $\phi=\pi/2$ or $\phi=-\pi/2$. Stability is defined by rotation frequency $\Omega$ as a function of coordinate $\phi$ at this point. Specific orientation $\phi_{0}$ is stable if $\partial\Omega /\partial t > 0$ in its vicinity. In this case direction of motion is directed towards $\phi_{0}$. Instability, therefore requires $\partial\Omega /\partial t > 0$. (B) Stability depend on position of rotation axis relative to the asymmetric part. Intuitively,  triangle is directed against the motion if it is dragged by a string (dashed arrow) connected to rotation axis near triangle's sharp corner. Triangle is directed together with the motion if the string is connected near the base of the triangle. Stability of a motor driven by thermal fluctuations requires more detailed analysis.}
\label{fig2}
\end{center}
\end{figure}

Analysis of motion emergence and stability requires at least two degrees of freedom, see Fig. \ref{fig2}. Motion will emerge if symmetric position of triangle $\phi=\pm\pi/2$ is unstable. This instability requires $\Omega$ as a function of $\phi$ near this point to be directed outward of the point. Opposite direction of $\Omega$ in the case of small deviations will return it to initial position. It is correct regarding the stability of maximum velocity at $\phi=0,\pi$.

In the case of two degrees of freedom $M,I>>m$, please see Supplementary Materials (SM) for detailed derivation, transition rate probability for each degree of freedom is similar to its single degree of freedom form (\ref{eq:30757}):
\begin{eqnarray}
  \label{eq:30969}
  &&W(V',\Delta V)=\frac{1}{4}\sum_{i}S_{i}\rho_{i}\sqrt{\frac{m}{2\pi T_{i}}}\\\nonumber
&&\left | \Delta V\right |H(\Delta V \Gamma_{V,i}(c))\int dcF(c)\Gamma_{V,i}^{2}(c)\left (\frac{M_{i}^{\star}(c)}{m\Gamma_{V,i}^{2}(c)}+1\right )^{2}\\\nonumber
&&\exp\left [-\frac{m\Gamma_{V,i}^{2}(c)\left ( V'+\frac{1}{2}\left[ \Delta V\left (\frac{M^{\star}_{i}(c)}{m\Gamma_{V,i}^{2}(c)}+1\right )\right ]\right )^{2}}{2T_{i}}\right ]
\label{eq:31254}    
\end{eqnarray}
with single difference that its mass $M(c)$ is a function of interaction with thermal bath. Transition to rotation occurs $V\rightarrow \Omega$, $M\rightarrow I$ and $\Gamma_{V}\rightarrow \Gamma_{\Omega}$. The rescaled mass and momentum are:
\begin{eqnarray}
  \label{eq:13}
  M^{\star}(c)=M(1+\frac{m}{I}\Gamma^{2}_{\Omega}(c)),I^{\star}(c)=I(1+\frac{m}{M}\Gamma^{2}_{V}(c)),
\end{eqnarray}
In the limit  $m/M,m/I\rightarrow 0$ expression (\ref{eq:30969}) converges to its single degree of freedom form (\ref{eq:30757}). Corrections (\ref{eq:13}), however, have significant impact on stability analysis of the motion.
 
Affect of multiple degrees of freedom on a single one $k$, similar to (\ref{eq:13}), reduces to rescaling of the mass in expression for transition rate probability (\ref{eq:30757}):
\begin{eqnarray}
  \label{eq:14}
  M_{k}^{\star}(c)=M_{k}(1+\sum_{j\neq k}\frac{m}{M_{j}}\Gamma^{2}_{j}(c)),
\end{eqnarray}
where sum goes over all degrees of freedom. This result is universal and stems from fluctuation dissipation theorem\cite{Kubo1966}, see SM.

Consider the case of two degrees of freedom (see Fig. \ref{fig1}), with rotation as a slow axis $m<<M<<I/r_{0}^{2}$, where $r_{0}$ is some characteristic radius. Rescaling of $M$ (\ref{eq:13}) introduces high order corrections to velocity $V$ (\ref{eq:3785554}) and, therefore, can be neglected. Translation motion, consequently, as in the case of single degree of freedom vanishes $V=0$ if $\phi=\pm\pi/2$ and takes maximum absolute value if $\phi=0,\pi$, see Fig. \ref{fig2}.

Rescaling of $I$ causes finite mean rotation $<\Omega>$ even if no rotation exists as a single degree of freedom:
\begin{eqnarray}
  \label{eq:2862}  <\Omega>_{\Gamma}=\frac{1}{2}\frac{\sqrt{m}}{M}\frac{S_{\bigtriangleup}\rho_{\bigtriangleup} T_{\bigtriangleup}^{1/2}\oint\Gamma_{\Omega,\bigtriangleup}\Gamma^{2}_{V,\bigtriangleup}}{\sum_{i=\bigtriangleup,\bigcirc}S_{i}\rho_{i} T_{i}^{1/2}\oint\Gamma^{2}_{\Omega,i}},
\end{eqnarray}
The next order if $\int\Gamma/G^{2}=0$ is:
\begin{eqnarray}
  \label{eq:28793}  <\Omega>_{\Gamma^{3}}=-\sqrt{\frac{\pi}{8}}\frac{m^{3/2}}{MI}\frac{S_{\bigtriangleup}\rho_{\bigtriangleup}(2T_{\bigtriangleup}-T^{eff}_{\Omega})\oint\Gamma^{3}_{\Omega,\bigtriangleup}\Gamma^{2}_{V,\bigtriangleup}}{\sum_{i=\bigtriangleup,\bigcirc}S_{i}\rho_{i} T_{i}^{1/2}\oint\Gamma^{2}_{\Omega,i}},
\end{eqnarray}
where $T^{eff}_{\Omega}$ is analogous to (\ref{eq:11}). Both expressions (\ref{eq:2862}) and (\ref{eq:28793}) vanish at points with $V=0$ $(\phi=\pm\pi/2)$ and $V=V_{max}$ $(\phi=0,\pi/2)$ due to symmetry. Near these points, therefore, rotation directed either outward of unstable point and convergences to stable points.

Stability depends on the sign of contour integral $\oint\Gamma_{\Omega}\Gamma_{V}^{2}$ in the case (\ref{eq:2862}), and  the sign of contour integral $\oint\Gamma_{\Omega}^{3}\Gamma_{V}^{2}$ together with the temperatures of the thermal bathes in the case (\ref{eq:28793}). The sign of the contour integrals depends on  the position of the rotation axis. Intuitively it can be understood as the different directions of triangle while being pulled by a string connected to rotation axis near its sharp node or base.

\begin{figure}
\resizebox{0.5\textwidth}{!}{\includegraphics{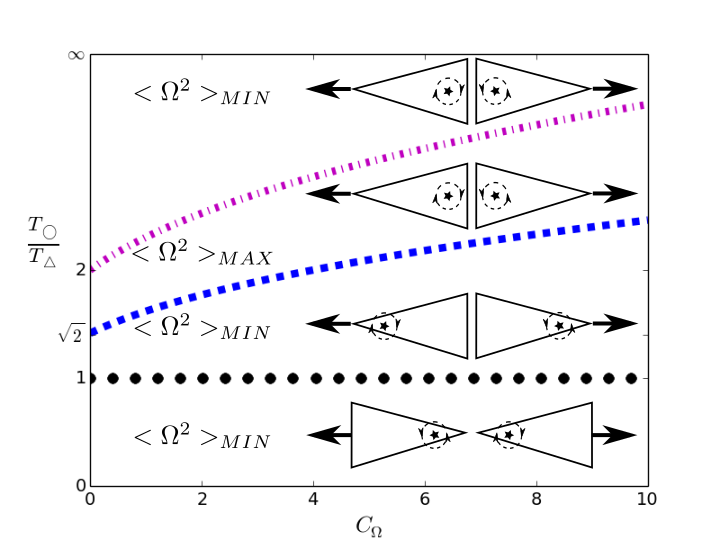}}
\caption{Stable direction of motion and of orientation in the space of a Brownian motor with two degrees of freedom. Stable configuration of the motor depends on the ratio  $T_{\bigcirc}/T_{\bigtriangleup}$ of the temperatures in the thermal bathes and on the ratio $C_{\Omega}$ of the gas densities multiplied by the size of the motor's parts in different thermal bathes. Stable motion occurs at four configurations. Direction of motion depends only on the ratio of the temperatures. Both temperatures and geometric factor affect stability of a motor with specific position of rotation axis. For instance, motor with rotation axis near its base can move only in the direction of its sharp corner. Fluctuations of rotation frequency take either maximum $<\Omega^{2}>_{MAX}$ or minimum $<\Omega^{2}>_{MIN}$ values at the stable configuration.}
\label{fig3}
\end{figure}

The stability and possible motion of the motor are presented in Fig. \ref{fig3} as a function of temperature ratio $T_{\bigcirc}/T_{\bigtriangleup}=1$ and parameter $c = \rho_{\bigtriangleup} S_{\bigtriangleup}/(\rho_{\bigcirc} S_{\bigcirc} )$. The boundary $T_{\bigcirc}/T_{\bigtriangleup}=1$ splits possible mutual directions of motion and asymmetry. This boundary is valid both for (\ref{eq:2862}) and (\ref{eq:28793}). Other boundaries define stability as a function of rotation axis position. These boundaries occur only in the case (\ref{eq:28793}). 

For each region $<\Omega^{2}>$ is indicated whether it takes maximum or minimum value at stable point. Near thermal equilibrium stability of rectified Brownian motion corresponds to minimum fluctuations $\Omega^{2}_{MIN}$ of the degree of freedom that defines stability of the motion. This condition however is not  universal, see Fig. \ref{fig3}.

Stability analysis of rectified motion (see Fig. \ref{fig3}) suggests break of chirality in the case (\ref{eq:28793}). Triangle with axis of rotation near its base can move only the direction of the asymmetry, such as direction of the sharp node of the triangle. Consequently, if a pool of the mechanical systems of \ref{fig1} type with arbitrary location of rotation axis evolves motion by developing a dissipation channel, e.g. ATP hydrolysis active cite, then motion directed together with spatial asymmetry of the system has an advantage. Chirality here corresponds to the relative directions of velocity and spatial asymmetry. Strictly speaking chirality appears if the triangle moves along a circle (rather than along the line) with either sharp node or base towards the direction of the motion.

This work, to the best of our knowledge, indicates the first thermodynamic explanation for chiral asymmetry in living matter. Chirality of living matter remains an open problem despite several proposals that include chiral symmetry break by strong external electromagnetic field\cite{Rikken2000}, by some asymmetric chemical agent (for review\cite{Guijarro2008,Barron2008}) and by different damage thresholds of L and R chiralities\cite{Dreiling2014,Rosenberg2008}. The strength of thermodynamic affect on chirality in real molecules remains to be determined. 

Extension of Onsager equations (\ref{eq:3}) to non linear regime is:
\begin{eqnarray}
  \label{eq:ytuyt456}
  \frac{\partial x}{\partial t} &=& L_{11}F_{x}+<V(\phi)>+L_{13}F_{\phi},\\\nonumber
  \frac{\partial Q}{\partial t} &=& L_{21}(\phi) F_{x}+L_{22}\frac{\Delta T}{T^{2}}+L_{23}F_{\phi},\\\nonumber
  \frac{\partial \phi}{\partial t} &=& L_{31}F_{x}+L_{32}\frac{\Delta T}{T^{2}}+L_{33}F_{\phi}+<\Omega(\phi)>,   
\end{eqnarray}
where near thermal equilibrium $<V(\phi)> \approx L_{12}(\phi)\Delta T/T^{2}$. According to this work, if degree of freedom $x$ depends on another one $\phi$ then $\phi$ acquires additional flux $<\Omega>$. 

Emergence of motion seems to require neither maximum\cite{England2013} nor minimum\cite{Prigogine1967} of dissipation $\partial Q/\partial t$, because according to (\ref{eq:ytuyt456}) it is independent of the flux $<\Omega(\phi)>$ that defines the stability of motion.

Hypothesis of motor emergence by transition between different mechanical states is supported by similarity between molecular motors and form changing signal proteins. This hypothesis, moreover, may explain the puzzle why these molecules possess a common ancestor\cite{Vale2000}.

Main limitations of the proposed stability analysis are the lack of the external potentials and the assumption of translation invariance of the degrees of freedom. For instance, these limitations prevent analysis of mechanical bodies that include springs. The main properties of mechanical Brownian ratchet is preserved in the limit of rigid connections(\cite{Parrondo1996,Sekimoto1997,Gomez-Marin2006}).

To conclude, this work demonstrates that mechanical Brownian motors can emerge by a universal thermo-dynamic mechanism in a fluctuating environment. For a broad class of the mechanical systems, the fluxes and their fluctuations universally depend on the geometric factors that define linear viscous coefficients. A novel method makes possible detailed modeling of an interaction between macroscopic body and multiple thermal bathes\cite{Gaspard2013,Zwanzig2001}.


\newpage
\section{Supplementary Material}
\setcounter{page}{1}
A. Feigel and A. Rozen, "Thermodynamic Emergence of a Brownian Motor".
\subsection{Calculation of velocity moments $<V^{n}>$}
Following\cite{Meurs2004}, here are the steps to calculate average moments of velocity $<V^{n}>$ in a 1D stochastic system. First, one derives transition rate probability $W(V,V')$ between velocities $V'\rightarrow V$. Second, probability $P(V,t)$ for velocity $V$ at time $t$ fits Boltzmann Master equation:
\begin{eqnarray}
  \label{eq:34}
  \frac{\partial P(V,t)}{\partial t} = &&\int W(V-\Delta V,\Delta V)P(V-\Delta V,t)d\Delta V - \\\nonumber
&&P(V,t)\int W(V,-\Delta V)d\Delta V,
\end{eqnarray}
where $\Delta V= V-V'$. Third, to calculate the moments:
\begin{eqnarray}
  \label{eq:45}
    <x^{n}> &=& \int x^{n}P(x,t)dx, x=\sqrt{\frac{M}{T^{eff}_{0}}}V,
\end{eqnarray}
eq. (\ref{eq:45}) is transformed using Kramers Moyal expansion:
\begin{eqnarray}
  \label{eq:34}
  \frac{\partial P(V,t)}{\partial t} = \sum_{n=1}^{\infty}\frac{(-1)^{n}}{n!}\frac{d^{n}}{dV^{n}}\left [a_{n}(V)P(V,t) \right ],
\end{eqnarray}
into a system of equations:
\begin{eqnarray}
  \label{eq:35}
  \partial <x>&=&<A_{1}(x)>,\\\nonumber
  \partial <x^{2}>&=&2<xA_{1}(x)>+<A_{2}(x)>,\\\nonumber
  \partial <x^{3}>&=&...,
\end{eqnarray}
where:

\begin{eqnarray}
  \label{eq:47}
  A_{n}&=&\left (\sqrt{\frac{M}{T^{eff}_{0}}} \right )^{n}a_{n},\\\nonumber
  a_{n}(V) &=& \int \Delta V^{n}W(V,\Delta V)d\Delta V,
\end{eqnarray}
Forth, the system (\ref{eq:35}) is solved for $<x^{n}>$ under assumption of steady state $\partial<x^{n}>=0$. At this stage additional expansion by some small parameter is required to linearize equations (\ref{eq:35}) in $<x^{n}>$.

\subsection{Calculation of transition rate probability $W$}
\subsubsection{Single degree of freedom - translation}

Here is calculation of transition rate probability $W(V,V')$ from velocity $V'$ to any other velocity $V$ for a macroscopic body of general 2D convex shape in an ideal gas\cite{Meurs2004}. The body possesses mass $M$ and can move along axis $x$ only. The gas consists of identical particles of mass $m$ at density $\rho$ and temperature $T$. The velocities of gas particles $(v_{x},v_{y})$ follow Maxwell distribution. In the case body is coupled to the several thermal bathes the final $W$ is the sum over the thermal bathes.

This work highlights universality of $W$ for different systems. It allows further extension of the method to the systems with multiple degrees of freedom. 

Consider collision of the body with a particle of the gas. In the frame of the reference of the body ($V'=0$), conservation of energy before and after the collision is:
\begin{eqnarray}
  \label{eq:9}
&&  -MV^2+\\\nonumber
&& m(v'_x-v_x)(v'_x+v_x)+m(v'_y-v_y)(v'_y+v_y)=0,
\end{eqnarray}
where $'$ indicates velocities before collision. Momentum conservation along x axis is:
\begin{eqnarray}
  \label{eq:mom2}
  -m\Delta v_x=MV,
\end{eqnarray}
where:
\begin{eqnarray}
  \label{eq:mom1}
  \Delta v_x=v_{x}-v'_{x},
\end{eqnarray}
In addition, in the case of instantaneous collision, momentum of the gas particle conserves along surface of body:
\begin{eqnarray}
  \label{eq:momsurf}
  v'_x\cos\theta+v'_y\sin\theta =  v_x\cos\theta+v_y\sin\theta,
\end{eqnarray}
where $\theta$ is the angle between surface of the body and axis $x$. In addition, it is convenient to express $\Delta v_{y}=v_{y}-v'_{y}$ as a function of $V$: 
\begin{eqnarray}
  \label{eq:21}
  \Delta v_{y}=\frac{M}{m}\frac{V}{\tan{\theta}}.
\end{eqnarray}
It follows from (\ref{eq:9}) and (\ref{eq:momsurf}) is:

An important consequence of the conservation laws (\ref{eq:mom2}) and (\ref{eq:21}) is:
\begin{eqnarray}
  \label{eq:28}
  \Delta v_{x}^{2}+\Delta v_{y}^{2}=\frac{M^{2}}{m^{2}}\frac{V^{2}}{\sin^{2}{\theta}},
\end{eqnarray}
In this expression we will see that geometric factor in the right part of this expression universally defines linear viscous coefficient for the corresponding degree of freedom.

To find velocity of the body $V$ after collision as a function of a gas particle velocity prior the collision $(v'_{x},v'_{y})$ let us rewrite energy conservation (\ref{eq:9}) as:
\begin{eqnarray}
  \label{eq:kin2}
&&-MV^2- m\Delta v_{x}^{2}-m\Delta v_{y}^{2}-\\\nonumber
&&2mv'_{x}\Delta v_{x}-2mv'_{y}\Delta v_{y}=0,
\end{eqnarray}
It can be rewritten further using (\ref{eq:mom2}) and expressing $\Delta v_{x}$ and $\Delta v_{y}$ as a functions of $V$:
\begin{eqnarray}
  \label{eq:kinlin1}
  -V-\frac{M}{m}\frac{V}{\Gamma^{2}_{V}(c)}+2v'_x-2v'_y\frac{1}{\tan\theta}=0,
\end{eqnarray}
where $\Gamma_{V}=\sin{\theta}$. The final formula is:
\begin{eqnarray}
  \label{eq:Vlinfin}
  V = \frac{2m/M\Gamma^{2}_{V}(c)}{1+m/M\Gamma^{2}_{V}(c)}\left [v'_x-v'_y\frac{1}{\tan\theta} \right ],
\end{eqnarray}
Transition from the frame of reference of the body to the laboratory frame of reference:
\begin{eqnarray}
  \label{eq:Vlinfin2}
  V = V'+\frac{2m/M\Gamma^{2}_{V}(c)}{1+m/M\Gamma^{2}_{V}(c)}\left [v'_x-V'-v'_y\frac{1}{\tan\theta} \right ],
\end{eqnarray}
requires addition of $-V'$ to the velocities along $x$ coordinate:

Taking into account (\ref{eq:Vlinfin2}), transition probability $V'\rightarrow V$ is:
\begin{eqnarray}
  \label{eq:tr1}
  &&dW(VV')=SF(\theta)d\theta\int_{-\infty}^{\infty} dv'_x\int_{-\infty}^{\infty}dv'_y\\\nonumber
&&H[(V'-v')e_\perp]\times (V'-v')e_\perp \rho \phi(v'_x,v'_y)\times\\\nonumber
&&\times\delta\left [  V -V'-\frac{2m/M\Gamma^{2}_{V}(c)}{1+m/M\Gamma^{2}_{V}(c)}\left [v'_x-V'-v'_y\frac{1}{\tan\theta} \right ] \right ], 
\end{eqnarray}
where $H$ is Heaviside step function and $\phi$ is Maxwell distribution:
\begin{eqnarray}
  \label{eq:max1}
  \phi(v'_{x},v'_{y})=\frac{m}{2\pi T}\exp\left (\frac{-m(v_{x}^{\prime 2}+v_{y}^{\prime 2})}{2T} \right ),
\end{eqnarray}
The average is calculated using Hadamard Stratonovich transformation.
The result is:
\begin{eqnarray}
  \label{eq:30}
  &&W(V,V')=\frac{1}{4}\sum_{i}S_{i}\rho_{i}\sqrt{\frac{m}{2\pi T_{i}}}\\\nonumber
&&\left | \Delta V\right |H(\Delta V \Gamma(c))\int dcF(c)\Gamma_{bath}^{2}(c)\left (\frac{M}{m\Gamma_{body}^{2}(c)}+1\right )^{2}\\\nonumber
&&\exp\left [-\frac{m\Gamma_{bath}^{2}(c)\left ( V'+\frac{1}{2}\left[ \Delta V\left (\frac{M}{m\Gamma_{body}^{2}(c)}+1\right )\right ]\right )^{2}}{2T_{i}}\right ],    
\end{eqnarray}
where $\Gamma_{body}\equiv\Gamma_{V}$ and  $\Gamma_{bath}=\Gamma_{body}$ to fit detailed balance in equilibrium.

\subsubsection{$\Gamma_{bath}=\Gamma_{body}$ equality from  detailed balance at thermal equilibrium}
Detailed balance at thermal equilibrium requires:
\begin{eqnarray}
  \label{eq:8}
  P^{eq}(V')W(V,V')=P^{eq}(-V)W(-V',-V),
\end{eqnarray}
where $P^{eq}(V)$ is distribution of velocities at thermal equilibrium. In the case of single translation degree of freedom:
\begin{eqnarray}
  \label{eq:29}
  P^{eq}(V)\propto\exp\left (-\frac{MV^{2}}{2T}\right ),
\end{eqnarray}
Detailed balance (\ref{eq:8}), taking into account (\ref{eq:30}) with $T_{i}=T_{j}$, reduces to:
\begin{eqnarray}
\nonumber
&&m\Gamma_{bath}^{2}\left (V'^{2}+V'(V-V')K+\frac{1}{4}(V-V')^{2}K^{2}\right )-\nonumber\\
&&m\Gamma_{bath}^{2}\left (V^{2}+V(V'-V)K+\frac{1}{4}(V-V')^{2}K^{2}\right )=\nonumber\\
&&MV'^{2}-MV^{2},
\label{eq:4}
\end{eqnarray}
where:
\begin{eqnarray}
  \label{eq:5}
  1-K=-\frac{M}{m\Gamma_{body}^{2}},
\end{eqnarray}
Finally, equality of geometric factors:
\begin{eqnarray}
  \label{eq:6}
  \Gamma^{2}_{bath}=\Gamma^{2}_{body},
\end{eqnarray}
follows from (\ref{eq:4}) and (\ref{eq:5}). 

\subsubsection{Single degree of freedom - rotation case}

Conservation of energy, conservation for angular momentum are:
\begin{eqnarray}
  \label{eq:rotsys1}
  -I\Omega^2+m(v'_x-v_x)(v'_x+v_x)+m(v'_y-v_y)(v'_y+v_y),
\end{eqnarray}
\begin{eqnarray}
  \label{eq:43}
    I\Omega+m\overrightarrow{r}\times \overrightarrow{v}=m\overrightarrow{r}\times
 \overrightarrow{v'},
\end{eqnarray}
Taking into account (\ref{eq:momsurf}) one gets:
\begin{eqnarray}
  \label{eq:38}
  \Delta v_{x}^{2}+\Delta v_{y}^{2}=\frac{I^{2}\Omega^{2}}{m^{2}(r_{x}\cos\theta+r_{y}\sin\theta)^{2}},
\end{eqnarray}

\begin{eqnarray}
  \label{eq:42} -\Omega-\frac{I}{m}\Omega\frac{1}{\Gamma^{2}_{\Omega}(c)}+\frac{2v'_{x}}{r_{y}+r_{x}/\tan\theta}-\frac{2v'_{y}}{r_{y}\tan\theta+r_{x}}=0,
\end{eqnarray}
where:
\begin{eqnarray}
  \label{eq:36}
  &&\Gamma_{\Omega}=\overrightarrow{r}\hat{e}_\parallel=\nonumber\\
  &&r_{x}\cos\theta+r_{y}\sin\theta,
\end{eqnarray}
and:
\begin{eqnarray}
  \label{eq:ev1755}
  &&\hat{e}_\perp=(\sin\theta,-\cos\theta),\\\nonumber
  &&\hat{e}_\parallel=(\cos\theta,\sin\theta),
\end{eqnarray}

Equation (\ref{eq:42}) is analogous to (\ref{eq:42}). Consequently $W(\Omega',\Omega)$ follows from (\ref{eq:30}) by substitution $V\rightarrow\Omega$, $M\rightarrow I$ and $\Gamma_{V}\rightarrow\Gamma_{\Omega}$.
\subsubsection{Two degrees of freedom}
In the case of both rotation and translation degrees of freedom, 
conservation laws become:
\begin{eqnarray}
  \label{eq:rotsys12}
  &&-MV^2-I\Omega^2+m(v'_x-v_x)(v'_x+v_x)+\\\nonumber
  &&m(v'_y-v_y)(v'_y+v_y)=0,\\\nonumber
  &&-v'_x\sin\phi+v'_y\cos\phi=-v_x\sin\phi+v_y\cos\phi,\\\nonumber
  &&-m(v'_x-v_x)=MV,\\\nonumber
  &&I\Omega+r\times v=r\times v',
\end{eqnarray}
In this case, the expressions (\ref{eq:28}) and (\ref{eq:38}) hold separately and imply: 
\begin{eqnarray}
  \label{eq:4076}
  M^{2}V^{2}\frac{1}{\Gamma_{V}^{2}(c)}=I^{2}\Omega^{2}\frac{1}{\Gamma_{\Omega}^{2}(c)},
\end{eqnarray}
Using (\ref{eq:4076}), energy conservation law:
\begin{eqnarray}
  \label{eq:41}
  -MV^{2}-I\Omega^{2}-m\sum_{i}\Delta v_{i}^{2}+2m\sum_{i}v_{i}\Delta v_{i}=0
\end{eqnarray}
may be rewritten as:
\begin{eqnarray}
  \label{eq:42} -V-\frac{M}{I}V\frac{\Gamma^{2}_{\Omega}(c)}{\Gamma^{2}_{V}(c)}-\frac{M}{m}V\frac{1}{\Gamma^{2}_{V}(c)}+\sum_{i}v_{i}f_{i}(c)=0,
\end{eqnarray}
and further reduced to:
\begin{eqnarray}
  \label{eq:42} -V-\frac{M}{m}V\frac{\mathfrak{G}^{2}_{V}(c)}{\Gamma^{2}_{V}(c)}+\sum_{i}v'_{i}f_{i}(c)=0,
\end{eqnarray}
The same can be done for rotation coordinate:
\begin{eqnarray}
  \label{eq:42} -\Omega-\frac{I}{m}\Omega\frac{\mathfrak{G}^{2}_{\Omega}}{\Gamma^{2}_{\Omega}(c)}+\sum_{i}v'_{i}g_{i}(c)=0,
\end{eqnarray}
\begin{eqnarray}
  \label{eq:53879}
  \mathfrak{G}^{2}_{V}=1+\frac{m}{I}\Gamma^{2}_{\Omega},
\end{eqnarray}
\begin{eqnarray}
  \label{eq:5365}
  \mathfrak{G}^{2}_{\Omega}=1+\frac{m}{M}\Gamma^{2}_{V},
\end{eqnarray}
Each degree of freedom possesses dynamics analogous to the single degree cases (\ref{eq:kinlin1}) and (\ref{eq:42}) with rescaled mass:
\begin{eqnarray}
  \label{eq:46}
  M(c) \rightarrow M\mathfrak{G}_{V}^{2},
\end{eqnarray}
and momentum:
\begin{eqnarray}
  \label{eq:49}
  I(c)\rightarrow I\mathfrak{G}_{\Omega}^{2},
\end{eqnarray}
correspondingly. The rescaling of $M$ and $I$ remains unaffected during derivation of transition rate probability (\ref{eq:30}) using independently (\ref{eq:42}) and (\ref{eq:42}). The requirement for detailed balance at thermal equilibrium also preserves. Consequently, final expression of $W$ for each degree of freedom take its single degree of freedom form (\ref{eq:30}) with rescaled $M$ (\ref{eq:46}) and $I$ (\ref{eq:49}), see (\ref{eq:30969}).

Transition to original frame of reference in the case of two degrees of freedom changes in (\ref{eq:Vlinfin2}) velocities $(v'_{x},v'_{y})$ to $(v'_{x}+S_{x},v'_{y}+S_{y})$, where $(S_{x},S_{y})$ are the velocity of the second degree of freedom. Transition rate probability (\ref{eq:tr1}) should be averaged over $(S_{x},S_{y})$. Distribution of $(S_{x},S_{y})$ is much more narrow than Maxwell distribution of the gas particles and absolute values $|(S_{x},S_{y})|<<|(v'_{x},v'_{y})|$ under assumption $m<<M$. In this case  $(S_{x},S_{y})$ can be neglected in derivation of transition rate probability ~(\ref{eq:30}). The same is valid in the case of arbitrary degrees of freedom.

\subsubsection{General N degree of freedom case}

Consider macroscopic body with arbitrary degrees of freedom that is immersed in an a gas of particles with mass $m$. As a consequence of the conservation laws, for each degree of freedom $k$ of the body holds:
\begin{eqnarray}
  \label{eq:40}
  \sum_{i}\Delta v_{i}^{2}=\frac{M_{k}^{2}}{m^{2}}V_{k}^{2}\frac{1}{\Gamma_{k}^{2}(c)},
\end{eqnarray}
where the sum goes over all degrees of freedom of the gas particles, e.g see expressions (\ref{eq:28}) and (\ref{eq:38}). The mass $M_{k}$ and velocity $v_{k}$ define energy of the corresponding degree of freedom $\propto M_{k}V_{k}^{2}$. For instance in the case of rotation mass $M$ together with velocity $V$ are replaced by momentum $I$ and frequency $\Omega$ correspondingly. Parameter $c$ in a geometric model indicates point of interaction on contour of the body but in general indicates channel of interaction with thermal bath.
 
As a consequence of (\ref{eq:40}), for each two degrees of freedom $k$ and $l$ holds:
\begin{eqnarray}
  \label{eq:407876}
  M_{k}^{2}V_{k}^{2}\frac{1}{\Gamma_{k}^{2}(c)}=M_{l}^{2}V_{l}^{2}\frac{1}{\Gamma_{l}^{2}(c)},
\end{eqnarray}
This expression connects different degrees of freedom.
  
The main task is to show that geometric factor $\Gamma (c)$ indeed defines linear viscous coefficient and therefore universally linked with macroscopic measurable parameter. In the limit of thermal equilibrium expression (\ref{eq:40}) should converge to diffusion coefficient of the body in velocity space. This diffusion coefficient is:
\begin{eqnarray}
  \label{eq:44}
  D_{V}=\left <\frac{V^{2}}{\Delta t}\right >=\frac{T\gamma}{M}
\end{eqnarray}
as a consequence of fluctuation dissipation theorem\cite{Kubo1966}. Thus  $\oint \Gamma^{2}\propto \gamma$ under assumption that $\sum \Delta v_{i}^{2}\propto T$ at thermal equilibrium. Indeed,  $\oint \Gamma^{2}\propto \gamma$ in the cases of the Brownian motors with either translation or rotation degrees of freedom.

These expressions (\ref{eq:40}) and (\ref{eq:407876}) are very general because correct number of constraint always allow reduction of conservation laws to (\ref{eq:40}). For instance, if the surface of the body is not convex then particle of the gas can make several collisions during the impact and expression (\ref{eq:momsurf}) should be different. The expression of the type (\ref{eq:40}) still can be derived and and it connection with viscosity is preserved. The same is correct for arbitrary number of degrees of freedom.
\begin{eqnarray}
  \label{eq:309837}
  &&W(V_{\xi},\Delta V_{\xi})=\frac{1}{4}\sum_{i}S_{i}\rho_{i}\sqrt{\frac{m}{2\pi T_{i}}}\int dc F_{i}(c)\\\nonumber
&&\left | \Delta V_{\xi}\right |H\left (\Delta V_{\xi} \frac{\Gamma_{\xi,i}(c)}{ \mathfrak{G}_{\xi,i}(c)}\right )\Gamma_{\xi,i}^{2}(c)\left (\frac{M_{\xi} \mathfrak{G}^{2}_{\xi,i}(c)}{m\Gamma_{\xi,i}^{2}(c)}+1\right )^{2}\\\nonumber
&&\exp\left [-\frac{m\Gamma_{\xi,i}^{2}(c)\left ( V_{\xi}+\frac{1}{2}\left[ \Delta V_{\xi}\left (\frac{M_{\xi} \mathfrak{G}^{2}_{\xi,i}(c)}{m\Gamma_{\xi,i}^{2}(c)}+1\right )\right ]\right )^{2}}{2T_{i}}\right ]
\end{eqnarray}
where rescaling parameter:
\begin{eqnarray}
  \label{eq:597965}
  \mathfrak{G}^{2}_{\xi,i}(c)=1+\sum_{\xi'\neq \xi}\frac{m}{M_{\xi',i}}\Gamma^{2}_{\xi',i}(c),
\end{eqnarray}
and index $i$ goes over all thermal bathes.

\subsection{Rectification and Stability}

\begin{widetext}
\begin{figure}
\begin{center}
    \begin{tabular}{c}
\multicolumn{1}{l}{{\bf\sf A}}\\
\resizebox{1.0\textwidth}{!}{\includegraphics{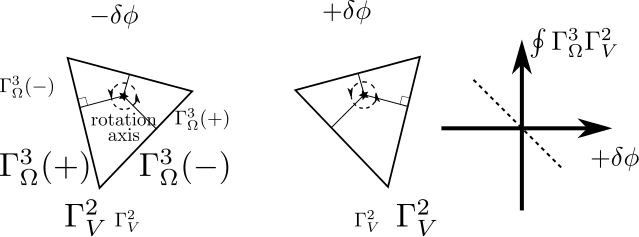}}\\
\multicolumn{1}{l}{{\bf\sf B}}\\
\resizebox{1.0\textwidth}{!}{\includegraphics{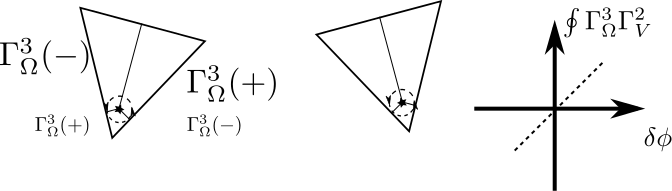}}\\
    \end{tabular}
\caption{Stability as a function of rotation axis position. (A) Axis of rotation is located near the base of the triangle. $\Gamma_{V}$ is constant along an edge of the triangle. Integral over the base is vanishes because of equal contribution of positive and negative $\Gamma_{\Omega}$. $\Gamma_{\Omega}$ changes sign at the intersection point of an edge and line perpendicular to the edge that crosses axis of rotation.  Indeed collision with particle on the different sides relative to this point will rotate triangle to opposite directions. The size of $\Gamma$ indicates absolute contribution to the contour integral either as a consequence of $\Gamma$ value or integration span along the surface. The sign of the contribution is provided inside the parenthesis. Contour integrals $\oint\Gamma_{\Omega}\Gamma^{2}_{V}$ and $\oint\Gamma^{3}_{\Omega}\Gamma^{2}_{V}$ possess negative derivative due to $\phi$ at the point $\phi=-\pi/2$. This point therefore is stable because flux $<\Omega>$ directed towards it. No motion can emerge. (B) Axis of rotation is located near the sharp corner. In this case, contour integrals $\oint\Gamma_{\Omega}\Gamma^{2}_{V}$ and $\oint\Gamma^{3}_{\Omega}\Gamma^{2}_{V}$ possess positive derivative due to $\phi$. The point is unstable and emergence of motion is possible.}
\label{fig4}
\end{center}
\end{figure}
\end{widetext}

To analyze rectification of motion and its stability, one should calculate the average velocities $<V_{\xi}>$  for the relevant degrees of freedom $\xi$. For instance, in the case  of a motor with two degrees of freedom, see Fig. \ref{fig1}, rectified motion $<V(\phi)>$ depends on orientation of the motor $\phi$. Consequently rotation $<\Omega(\phi)>$ defines stability of the orientations. Stable points are characterized by in flux of  $<\Omega(\phi)>$, see Figs. \ref{fig2} and \ref{fig4}.

To calculate the moments $<V_{\xi}^{n}>$, Boltzmann equation (\ref{eq:34}) with transition rate probability (\ref{eq:309837}) are transformed by Kramers Moyal expansion (\ref{eq:35}). In addition equations are expanded using small parameter $m/M_{\xi}$. The first three moments then are:
\begin{eqnarray}
  \label{eq:52}
&&\frac{\partial <V>}{\partial t} = \sum_{i}\rho_{i}S_{i}\sqrt{\frac{T_{i}}{m}}\\\nonumber
&&\left [ -\sqrt{\frac{T_{i}}{M}}\oint\frac{\Gamma_{V,i}(c)}{\mathfrak{G}^{2}_{V,i}(c)}\epsilon_{V}^{1}\right .\\\nonumber
&&-2\sqrt{\frac{2}{\pi}}\oint \frac{\Gamma_{V,i}^{2}(c)}{\mathfrak{G}_{V,i}^{2}(c)}<V>\epsilon_{V}^{2}\\\nonumber
&&\left . +\left ( \oint\frac{\Gamma_{V,i}^{3}}{\mathfrak{G}_{V,i}^{4}}\sqrt{\frac{T_{i}}{M}}-\oint\frac{\Gamma_{V,i}^{3}}{\mathfrak{G}_{V,i}^{2}}\sqrt{\frac{M}{T_{i}}}<V^{2}>\right ) \epsilon_{V}^{3} \right ],
\end{eqnarray}

\begin{eqnarray}
  \label{eq:524534}
&&\frac{\partial <V^{2}>}{\partial t} = \\\nonumber
&&\sum_{i} \rho_{i}S_{i}\sqrt{\frac{T_{i}}{m}} \left [ -2\sqrt{\frac{T_{i}}{M}}\oint\frac{\Gamma_{V,i}}{\mathfrak{G}^{2}_{V,i}}<V>\epsilon_{V}^{1}\right .\\\nonumber
&&\left .+4\sqrt{\frac{2}{\pi}}\left( \oint\frac{\Gamma_{V,i}^{2}}{\mathfrak{G}_{V,i}^{4}}\frac{T_{i}}{M}-\oint\frac{\Gamma_{V,i}^{2}}{\mathfrak{G}^{2}_{V,i}}<V^{2}>\right )\epsilon_{V}^{2} \right .\\\nonumber
&&\left . -2\left(-4 \oint\frac{\Gamma_{V,i}^{3}}{\mathfrak{G}_{V,i}^{4}}\sqrt{\frac{T_{i}}{M}}<V>+\sqrt{\frac{M}{T_{i}}}\oint\frac{\Gamma_{V,i}^{3}}{\mathfrak{G}^{2}_{V,i}}<V^{3}>\right )\epsilon_{V}^{3} \right ],
\end{eqnarray}
\begin{eqnarray}
  \label{eq:52479}
&&\frac{\partial <V^{3}>}{\partial t} = \\\nonumber
&&\sum_{i} \rho_{i}S_{i}\sqrt{\frac{T_{i}}{m}} \left [ -3\sqrt{\frac{T_{i}}{M}}\oint\frac{\Gamma_{V,i}}{\mathfrak{G}^{2}_{V,i}}<V^{2}>\epsilon_{V}^{1}\right .\\\nonumber
&&\left .+6\sqrt{\frac{2}{\pi}}\left( 2\oint\frac{\Gamma_{V,i}^{2}}{\mathfrak{G}_{V,i}^{4}}\frac{T_{i}}{\sqrt{M}}<V>-\sqrt{M}\oint\frac{\Gamma_{V,i}^{2}}{\mathfrak{G}^{2}_{V,i}}<V^{3}>\right )\epsilon_{V}^{2} \right ],
\end{eqnarray}
where $\epsilon_{V}=m/M$, index $i$ goes over the thermal bathes and the index $\xi$ is omitted.

Eqs. (\ref{eq:52}),~(\ref{eq:524534}) and (\ref{eq:52479}) converge to the corresponding results in the case of a system with single degree of freedom, if there is no affect of the other degrees of freedom $\mathfrak{G}=1$. In this case the terms with the first degree of $\Gamma$ vanish because $\oint\Gamma=0$.   

In the case of two degrees of freedom $V$ and $\Omega$:
\begin{eqnarray}
  \label{eq:53}
  \mathfrak{G}^{2}_{V,i}=1+\frac{m}{I}\Gamma^{2}_{\Omega,i},
\end{eqnarray}
\begin{eqnarray}
  \label{eq:536567}
  \mathfrak{G}^{2}_{\Omega,i}=1+\frac{m}{M}\Gamma^{2}_{V,i},
\end{eqnarray}
In the limit $m<<M,m<<I$:
\begin{eqnarray}
  \label{eq:59292}
  \frac{1}{\mathfrak{G}^{2}_{V,i}}\approx 1-\frac{m}{I}\Gamma^{2}_{\Omega,i},
\end{eqnarray}
\begin{eqnarray}
  \label{eq:511390}
  \frac{1}{\mathfrak{G}^{4}_{V,i}}\approx 1-2\frac{m}{I}\Gamma^{2}_{\Omega,i},
\end{eqnarray}
Equations for $\Omega$ follow by change $V\rightarrow\Omega$, $M\rightarrow I$, $\Gamma_{V}\rightarrow\Gamma_{\Omega}$ and $\epsilon_{V}\rightarrow\epsilon_{\Omega}$, where $\epsilon_{\Omega}=m/I$.

Neglecting the time derivatives in  (\ref{eq:52}),~(\ref{eq:524534}) and (\ref{eq:52479}) one gets:
\begin{eqnarray}
  \label{eq:10}  &&<V^{2}>=\frac{1}{M}\frac{\sum_{i}\rho_{i}S_{i}T_{i}^{\frac{3}{2}}\frac{\Gamma_{V,i}^{2}}{\mathfrak{G}_{V,i}^{4}}}{\sum_{i}\rho_{i}S_{i}T^{\frac{1}{2}}\frac{\Gamma_{V,i}^{2}}{\mathfrak{G}_{V,i}^{2}}}\\\nonumber
&&-\frac{1}{2\sqrt{M}}\sqrt{\frac{\pi}{2}}\frac{\sum_{i}\rho_{i}S_{i}T_{i}\oint\frac{\Gamma_{V,i}(c)}{\mathfrak{G}^{2}_{i}(c)}}{\sum_{i}\rho_{i}S_{i}T^{\frac{1}{2}}_{i}\oint \frac{\Gamma_{V,i}^{2}(c)}{\mathfrak{G}_{V,i}^{2}(c)}}\frac{<V>}{\epsilon_{V}}+\frac{1}{2}\sqrt{\frac{\pi}{2}}\times\\\nonumber
&&\frac{ \sum_{i}\rho_{i}S_{i}\left (4 T_{i}<V>\oint\frac{\Gamma_{V,i}^{3}}{\mathfrak{G}_{V,i}^{4}}-M<V^{3}>\oint\frac{\Gamma_{V,i}^{3}}{\mathfrak{G}_{V,i}^{2}}\right )}{ \sum_{i}\rho_{i}S_{i}T^{\frac{1}{2}}_{i}\oint \frac{\Gamma_{V,i}^{2}(c)}{\mathfrak{G}_{V,i}^{2}(c)}
 }\frac{\epsilon_{V}}{\sqrt{M}},
\end{eqnarray}
For the further calculations it is convenient to define the leading term as:
\begin{eqnarray}
  \label{eq:16} V^{2}_{0}=\frac{1}{M}\frac{\sum_{i}\rho_{i}S_{i}T_{i}^{\frac{3}{2}}\frac{\Gamma_{V,i}^{2}}{\mathfrak{G}_{V,i}^{4}}}{\sum_{i}\rho_{i}S_{i}T^{\frac{1}{2}}\frac{\Gamma_{V,i}^{2}}{\mathfrak{G}_{V,i}^{2}}},
\end{eqnarray}
The third moment is:
\begin{eqnarray}
  \label{eq:10760}  <V^{3}>=<V>\frac{2\sum_{i}\rho_{i}S_{i}T^{3/2}_{i}\oint\frac{\Gamma^{2}_{V,i}}{\mathfrak{G}_{V,i}^{4}}}{M\sum_{i}\rho_{i}S_{i}\sqrt{T_{i}}\oint\frac{\Gamma^{2}_{V,i}}{\mathfrak{G}_{V,i}^{2}}},
\end{eqnarray}
It can be rewritten as:
\begin{eqnarray}
  \label{eq:17}
  <V^{3}>=2<V>V^{2}_{0}.
\end{eqnarray}
using (\ref{eq:16}).

The leading term of velocity is:
\begin{eqnarray}
  \label{eq:1}
<V>_{\Gamma} = -\frac{1}{2\sqrt{M}}\sqrt{\frac{\pi}{2}}\frac{\sum_{i}\rho_{i}S_{i}T_{i}\oint\frac{\Gamma_{V,i}(c)}{\mathfrak{G}^{2}_{i}(c)}}{\sum_{i}\rho_{i}S_{i}T^{\frac{1}{2}}_{i}\oint \frac{\Gamma_{V,i}^{2}(c)}{\mathfrak{G}_{V,i}^{2}(c)}}\epsilon_{V}^{-1},  
\end{eqnarray}
It is of the order $\epsilon_{\Omega}^{2}/\epsilon_{V}$ taking into account (\ref{eq:59292}) and $\oint\Gamma=0$.
The corresponding second moment is:
\begin{eqnarray}
  \label{eq:19}
  <V^{2}>_{\Gamma}=V^{2}_{0}+<V>_{\Gamma}^{2},
\end{eqnarray}
In the case $\int\Gamma/G^{2}=0$:
\begin{eqnarray}
  \label{eq:2}
&&<V>_{\Gamma^{3}} =\\\nonumber
&&\frac{1}{2}\sqrt{\frac{\pi}{2}}\frac{ \sum_{i}\rho_{i}S_{i}\left ( T_{i}\oint\frac{\Gamma_{V,i}^{3}}{\mathfrak{G}_{V,i}^{4}}-MV_{0}^{2}\oint\frac{\Gamma_{V,i}^{3}}{\mathfrak{G}_{V,i}^{2}}\right )}{ \sum_{i}\rho_{i}S_{i}T^{\frac{1}{2}}_{i}\oint \frac{\Gamma_{V,i}^{2}(c)}{\mathfrak{G}_{V,i}^{2}(c)}
 }\frac{\epsilon_{V}}{\sqrt{M}},
\end{eqnarray}
and:
\begin{eqnarray}
  \label{eq:20}
&&<V^{2}>_{\Gamma^{3}} =V_{0}^{2}+\\\nonumber
&&\sqrt{\frac{\pi}{2}}\frac{ \sum_{i}\rho_{i}S_{i}\left (2 T_{i}\oint\frac{\Gamma_{V,i}^{3}}{\mathfrak{G}_{V,i}^{4}}-MV_{0}^{2}\oint\frac{\Gamma_{V,i}^{3}}{\mathfrak{G}_{V,i}^{2}}\right )}{ \sum_{i}\rho_{i}S_{i}T^{\frac{1}{2}}_{i}\oint \frac{\Gamma_{V,i}^{2}(c)}{\mathfrak{G}_{V,i}^{2}(c)}
 }\frac{<V>_{\Gamma^{3}}\epsilon_{V}}{\sqrt{M}},  
\end{eqnarray}
Expressions (\ref{eq:1}),~(\ref{eq:19}),~(\ref{eq:2}) and (\ref{eq:20}) are the main results of this work that describe mutual influence of degrees of freedom on each other.

In the case of two degrees of freedom and two thermal bathes, see Fig. \ref{fig1}:
\begin{eqnarray}
  \label{eq:22}
  <\Omega>_{\Gamma}=\frac{1}{2}\frac{\sqrt{m}}{M}\frac{S_{\bigtriangleup}\rho_{\bigtriangleup} T_{\bigtriangleup}^{1/2}\oint\Gamma_{\Omega,\bigtriangleup}\Gamma^{2}_{V,\bigtriangleup}}{\sum_{i=\bigtriangleup,\bigcirc}S_{i}\rho_{i} T_{i}^{1/2}\oint\Gamma^{2}_{\Omega,i}},
\end{eqnarray}
and:
\begin{eqnarray}
  \label{eq:23}  <\Omega>_{\Gamma^{3}}=-\sqrt{\frac{\pi}{8}}\frac{m^{3/2}}{MI}\frac{S_{\bigtriangleup}\rho_{\bigtriangleup}(2T_{\bigtriangleup}-T^{eff}_{\Omega})\oint\Gamma^{3}_{\Omega,\bigtriangleup}\Gamma^{2}_{V,\bigtriangleup}}{\sum_{i=\bigtriangleup,\bigcirc}S_{i}\rho_{i} T_{i}^{1/2}\oint\Gamma^{2}_{\Omega,i}}. 
\end{eqnarray}
Both (\ref{eq:22}) and (\ref{eq:23}) depend on $\phi$ because contour integrals with $\Gamma$ depend on $\phi$. 

Stability of specific orientation $\phi_{0}$ is defined by linear expansion of (\ref{eq:22}) or (\ref{eq:23}) in $\phi$ near $\phi_{0}$, see Figs. \ref{fig2} and \ref{fig4}. Stability in the case $\int\Gamma/G^{2}\neq 0$ depends only on the contour integral $\oint\Gamma_{\Omega,\bigtriangleup}\Gamma^{2}_{V,\bigtriangleup}$, because the temperature term in (\ref{eq:22}) is always positive. Orientation of the asymmetric part in this case is independent of the temperatures. If $\int\Gamma/G^{2}=0$ then: 
\begin{eqnarray}
  \label{eq:26}
  <\Omega>_{\Gamma^{3}}\propto -(2T_{\bigtriangleup}-T^{eff}_{\Omega})\oint\Gamma^{3}_{\Omega,\bigtriangleup}\Gamma^{2}_{V,\bigtriangleup},
\end{eqnarray}
and, therefore, stability depends on the temperatures of the thermal bathes.

Fluctuations of the velocities (\ref{eq:20}) are:
\begin{eqnarray}
  \label{eq:24}
  <\Omega^{2}>_{\Gamma}=\Omega_{0}^{2}+<\Omega>^{2}_{\Gamma},
\end{eqnarray}
and
\begin{eqnarray}
  \label{eq:25} 
&&<\Omega^{2}>_{\Gamma^{3}}=\Omega_{0}^{2}-\\\nonumber
&&\sqrt{\frac{\pi}{2}}\sqrt{\frac{m}{MI}}\frac{S_{\bigtriangleup}\rho_{\bigtriangleup}(4T_{\bigtriangleup}-T^{eff}_{\Omega})<\Omega>_{\Gamma^{3}}\oint\Gamma^{3}_{\Omega,\bigtriangleup}\Gamma^{2}_{V,\bigtriangleup}}{\sum_{i=\bigtriangleup,\bigcirc}S_{i}\rho_{i}T_{i}^{1/2}\oint\Gamma^{2}_{\Omega,i}},
\end{eqnarray}
correspondingly. In this case (\ref{eq:24}), any stable point with $<\Omega>_{\Gamma}=0$ correspond to minimum fluctuations. In the case of (\ref{eq:25}):
\begin{eqnarray}
  \label{eq:27} 
&&<\Omega^{2}>_{\Gamma^{3}}=\Omega_{0}^{2}+\\\nonumber
&&A_{+}(4T_{\bigtriangleup}-T^{eff}_{\Omega})(2T_{\bigtriangleup}-T^{eff}_{\Omega})\left (\oint\Gamma^{3}_{\Omega,\bigtriangleup}\Gamma^{2}_{V,\bigtriangleup}\right )^{2},  
\end{eqnarray}
where $A_{+}$ is a positive coefficient. According to (\ref{eq:27}), a stable point might correspond to minimum or maximum fluctuations as a function of the temperature of the thermal bathes.

Stability conditions $<\Omega>$ together with properties of $<\Omega^{2}>$ change at the temperature boundaries $(2T_{\bigtriangleup}-T^{eff}_{\Omega})=$ and $(4T_{\bigtriangleup}-T^{eff}_{\Omega})=0$, see Fig. \ref{fig3}. These equations reduce to:
\begin{eqnarray}
  \label{eq:7}
  c+2\frac{T_{\bigcirc}}{T_{\bigtriangleup}}-\left (\frac{T_{\bigcirc}}{T_{\bigtriangleup}}\right )^{3}=0,
\end{eqnarray}
and
\begin{eqnarray}
  \label{eq:12}
  3c+4\frac{T_{\bigcirc}}{T_{\bigtriangleup}}-\left (\frac{T_{\bigcirc}}{T_{\bigtriangleup}}\right )^{3}=0,  
\end{eqnarray}
correspondingly. Coefficient $c$ is:
\begin{eqnarray}
  \label{eq:15}
  c = \frac{\rho_{\bigtriangleup} S_{\bigtriangleup} }{\rho_{\bigcirc} S_{\bigcirc} },
\end{eqnarray}
\onecolumngrid
\end{document}